\documentclass[doublecol]{epl2} 
\def\ltsima{$\; \buildrel < \over \sim \;$}
\def\ltsim{\lower.5ex\hbox{\ltsima}}
\title{Jamming mechanisms and density dependence in a kinetically-constrained model}
\author{Yair Shokef\inst{1} \and Andrea J. Liu\inst{2}}
\shortauthor{Y. Shokef and A.J. Liu}
\institute{                    
  \inst{1} Department of Physics of Complex Systems, Weizmann Institute of Science, Rehovot 76100, Israel\\
  \inst{2} Department of Physics and Astronomy, University of Pennsylvania, Philadelphia, PA 19104, USA
}
\pacs{64.70.Q-}{Theory and modeling of the glass transition}
\pacs{45.70.-n}{Granular systems}
\pacs{47.57.-s}{Complex fluids and colloidal systems}

\abstract{
We add relaxation mechanisms that mimic the effect of temperature and non-equilibrium driving to the recently-proposed spiral model which jams at a critical density $\rho_c<1$. This enables us to explore unjamming by temperature or driving at $\rho_c<\rho<1$. We numerically calculate the relaxation time of the persistence function and its spatial heterogeneity.  We disentangle the three different relaxation mechanisms responsible for unjamming when varying density, temperature, and driving strength, respectively. We show that the spatial scale of dynamic heterogeneity depends on density much more strongly than on temperature and driving.
}

\begin{document}

\maketitle

\section{Introduction}

\begin{figure}
\onefigure[width=0.63\columnwidth,angle=270]{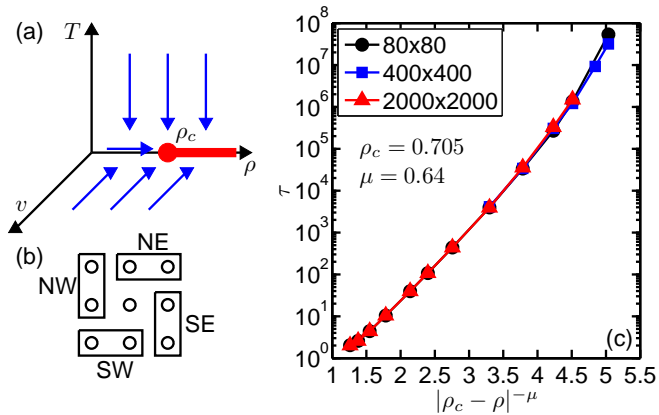}
\caption{(Color online) a) Jamming phase-diagram of density $\rho$, temperature $T$, and driving $v$. Jammed phase is the thick red line along the $\rho$-axis which terminates at $\rho_c$. Blue arrows are the trajectories we investigate. b) 
The spiral model is defined by dividing the neighbors of each site on the square lattice into four pairs, labeled NE, SE, SW, and NW (see text). c) Divergence of relaxation time as $\rho \rightarrow \rho_c$ for $T=0$ and $v=0$, with lattice size indicated in the legend.}
\label{fig:T0F0}
\end{figure}

Glass-forming-liquids, colloids, emulsions, foams, and granular matter all develop sluggish and heterogeneous dynamics as they approach the onset of jamming.  The slowing down of the dynamics in these systems with increasing density of the constituent particles, decreasing temperature, or decreasing the strength of external driving forces is often summarized in the form of a jamming phase-diagram~\cite{LiuNagel1998}.   To date, most numerical studies of this diagram have focused on particulate models such as sphere packings.  From a theoretical point of view, however, simpler models are easier to understand.  Here, we introduce a lattice model with a phase-diagram (fig.~\ref{fig:T0F0}a) that is similar to that of sphere packings, and use it to study dynamic heterogeneities.

Experiments on granular~\cite{AbateDurian2007} and colloidal~\cite{Brambilla} systems show steady growth in dynamic heterogeneities as the relaxation time increases with increasing density.  In glass-forming liquids, however, the scale of heterogeneities remains modest even as the relaxation time increases by more than 10 orders of magnitude with decreasing temperature~\cite{DalleFerrier}.  This difference may be due to the far greater dynamic range measurable in glass-forming liquids~\cite{Brambilla}.  Our model, however, suggests that this difference signals a fundamental distinction between jamming due to density as opposed to jamming by temperature or driving.

At zero temperature and driving, sphere packings undergo a jamming transition with density~\cite{Durian1995,OHern2002,OHern2003}. This transition has a mixed nature in that the number of interacting neighbors per sphere jumps discontinuously from zero to the minimum number needed for mechanical stability, but there are diverging length scales~\cite{OHern2002,OHern2003,Silbert}. Above the critical density, the spheres may be unjammed either by raising temperature above a glass transition into an equilibrium state or by applying shear stress above a yield stress to drive the system into a homogeneous non-equilibrium steady-state.  

\section{Model}

To construct a lattice model with a similar phase-diagram, we start with a kinetically-constrained lattice-gas.  In such models, occupied sites do not interact but the dynamical rules governing changes in occupation of a site depend on the occupation of neighboring sites.  This leads to dynamics that are increasingly slow and heterogeneous as the fraction of occupied sites increases, since increasingly larger regions are required to rearrange collectively~\cite{FA,KA,KCM_review}. In such models, the fraction of occupied sites might either be interpreted as a density variable or (in spin versions of the model) controlled by temperature.  Thus, temperature and density are equivalent variables in such models, and cannot be controlled independently of each other.  In order to generalize to the case where temperature and density are \emph{independent control variables},  we choose to associate the fraction of occupied sites with density, $\rho$, so that a conventional kinetically-constrained model corresponds to our zero-temperature model.  Here we focus on the spiral model~\cite{JSP_130,BiroliToninelli2008}, defined on the square lattice such that the occupation of a site can only change if its (NE or SW) and (NW or SE) neighboring pairs (see fig.~\ref{fig:T0F0}b) are completely empty. This model jams into a non-ergodic phase at $\rho_c \approx 0.705$, which allows us to study its behavior at $\rho_c<\rho<1$ with nonzero temperature or driving.  Such jamming-percolation models~\cite{ToninelliBiroliFisher2006,JengSchwarz_comment,Toninelli_reply
,JengSchwarz2008} possess an additional important property: the jamming transition at $\rho_c$ has a mixed nature; the fraction of stuck particles that cannot participate in rearrangements jumps discontinuously as in a first-order transition, while time and length scales diverge as in a second-order transition.  For a recent investigation of the equilibrium and non-equilibrium dynamics of the spiral model, see \cite{Corberi}.

In the original spiral model the number of particles was not conserved.  Following \cite{JSP_126}, we modify the zero-temperature stochastic dynamics so that instead of switching sites between being occupied and vacant we move a particle to a neighboring site if the target site is vacant and if the kinetic constraint described in fig.~\ref{fig:T0F0}b holds both before and after the move.  We measure time in units of attempted moves per particle.

We introduce temperature by softening the kinetic constraints.  Instead of preventing blocked moves, we allow them with probability $\exp(-1/T)$.  Thus, for $T=0$ we recover the original model with rigid constraints.  Note that the system still has no interactions and that energy is only associated with the virtual barrier the system has to cross in a kinetically-constrained move.  For $\rho>\rho_c$, the system is non-ergodic only at $T=0$, since for arbitrarily low temperature, kinetically-constrained moves occur at a slow but nonzero rate, hence the system may eventually reach any configuration.

We drive the system into a non-equilibrium steady-state by inducing a current of particles in one direction (for example, from left to right), as follows.  In addition to the afore-mentioned moves in which particles can move into neighboring vacant sites subject to the soft ($T>0$) or rigid ($T=0$) kinetic constraint, we introduce a second type of move, in which a particle can move into the neighboring site to its right if it is vacant, irrespective of the kinetic constraints~\cite{footnote_Fielding_Sellitto}.  Such moves are attempted at rate $f$, and we characterize the driving strength by the average flow velocity $v=(1-\rho)f$ induced by them.  Note that even for arbitrarily slow driving, the environment of any blocked particle will eventually change such that the particle will no longer be blocked and can move even under the kinetically-constrained dynamics~\cite{yield_stress_footnote}.

We could have chosen to soften the kinetic constraints in more complicated ways that couple density to temperature or driving to obtain more realistic results.  The advantage of our implementation is that the interplay of density, temperature and driving appears in its purest, simplest form.

We explored the model with rejection-free Monte-Carlo simulations, and established convergence of the results with system size by comparing systems of 80x80, 400x400, and 2000x2000 sites. To extract the relaxation time and the scale of dynamic heterogeneities, we calculate the persistence function, $p_i(t)$, defined as the probability that particle $i$ has not moved over a time interval $t$~\cite{footnote_px}. At long waiting times we find either exponential or stretched exponential decay, depending on density, of the particle-averaged persistence function, $p(t) \equiv 1/N \sum_i p_i(t)$, where $N$ is the number of particles in the system; we extract the relaxation time $\tau$ from the condition $p(\tau) = 1/e$.   We measure the dynamic heterogeneity in the standard way~\cite{Glotzer,Franz,Dauchot,Keys}, by calculating the variance of the persistence function 
\begin{equation}
\chi_4(t) = N [\langle p^2(t) \rangle - \langle p(t) \rangle ^2 ], 
\end{equation}
where $\langle \: \rangle$ denotes an average over different stochastic realizations for the initial state and dynamics.   This quantity has been shown to reflect the spatial extent of dynamic heterogeneities~\cite{AbateDurian2007,Toninelli_PRE_2005}. The idea behind this is that if relaxation is heterogeneous, different regions in space, or alternatively different copies of the system, relax at different times, hence there is a large variance in $p(t)$ between them.

In the absence of temperature or driving, relaxation slows down with increasing density not only because each particle is more likely to be blocked by its neighbors but also because these neighbors are in turn blocked by their neighbors and so on. Hence moving a particle requires a collective motion of many other particles. In the spiral model, as $\rho\rightarrow\rho_c$, the size of blocking clusters is expected to diverge as $\log (L)\sim\left(\rho_c-\rho\right)^{-\mu}$ with $\mu = 0.64$~\cite{BiroliToninelli2008,ToninelliBiroliFisher2006}.  Fig.~\ref{fig:T0F0}c shows that we find $\log(\tau)\sim\left(\rho_c-\rho\right)^{-\mu}$.  This is consistent with a scaling relation between length scales and time scales of the form $\tau \sim L^z$,  so that $\log(\tau) \sim \log(L)$, irrespective of the scaling exponent $z$.

\begin{figure}
\onefigure[width=\columnwidth]{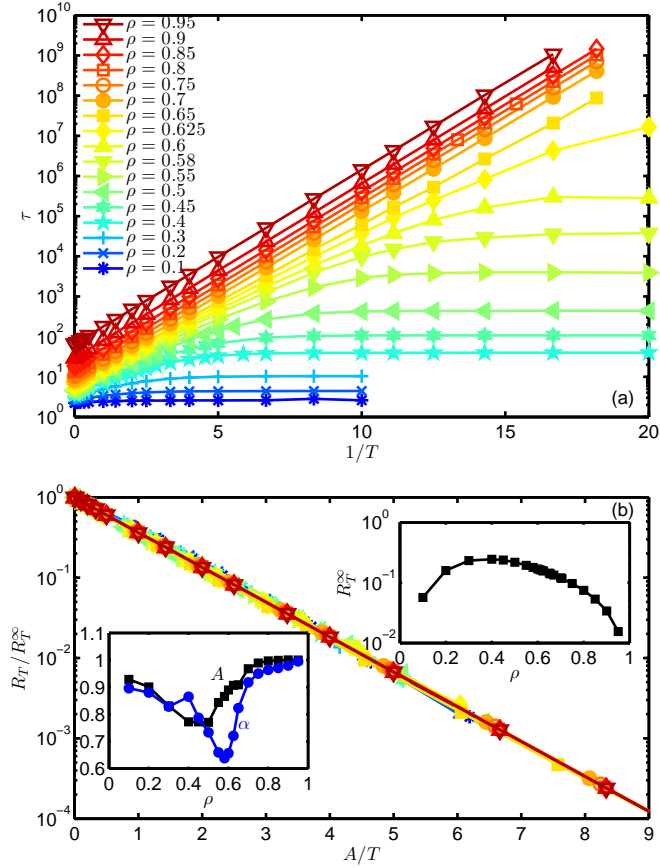}
\caption{(Color online) a) Relaxation time vs inverse temperature for various densities. b) Relaxation rate due to the temperature mechanism at all densities may be collapsed to Arrhenius form by normalizing $R_T$ by its $T=\infty$ value given in the top inset, and scaling $T$ by the effective barrier height $A(\rho)$ given in the bottom inset.}
\label{fig:T}
\end{figure}

\section{Jamming Mechanisms}

The behavior of the relaxation time as a function of temperature and density at zero driving is summarized in fig.~\ref{fig:T}a.  As $T \rightarrow 0$, $\tau$ diverges for $\rho>\rho_c$, while for $\rho<\rho_c$ it saturates to the finite $T=0$ value given in fig.~\ref{fig:T0F0}c.  The singularity at $\rho_c$ and $T=0$ affects the behavior of $\tau(\rho,T)$ nearby.  However, the overall relaxation rate we measure arises from a combination of two types of physical processes, which may be attributed to density and temperature separately. We demonstrate this by writing the relaxation rate as 
\begin{equation}
\frac{1}{\tau(\rho,T)} = R_{\rho}(\rho) + R_T(\rho,T).
\end{equation}
Here, $R_{\rho} \equiv 1/\tau(\rho,T=0)$ is the relaxation rate due to the \emph{density mechanism}, which represents processes subject to the kinetic constraints in which the neighborhood of a particle changes so that a previously-blocked particle can move.  Thus, $R_\rho=0$ for $\rho \ge \rho_c$. Similarly, $R_T$ is the relaxation rate due to the \emph{temperature mechanism}, representing the process in which a blocked particle moves either by directly overcoming the kinetic constraint or by becoming unblocked when one of its neighbors overcomes its kinetic constraint by thermal activation.  Clearly, 
\begin{equation}
R_T = \frac{1}{\tau(\rho,T)} - \frac{1}{\tau(\rho,T=0)}.  
\end{equation}

Figure~\ref{fig:T}b shows that $R_T$ has an Arrhenius dependence on temperature: $R_T = R_T^{\infty} \exp(-A(\rho)/T)$.  Here, $R_T^{\infty}(\rho)$ is the relaxation rate at $T=\infty$, where the kinetic constraint becomes irrelevant. For $\rho > \rho_c$, the dominant process facilitated by thermal activation is motion of a single blocked particle moving at probability $\exp(-1/T)$, therefore $A \approx 1$. For $\rho \ltsim \rho_c$, on the other hand, the density and temperature mechanisms are inherently coupled since moves are typically blocked by a large cluster of neighbors, and there are multiple moves that can lead to unblocking a single particle, hence $A<1$. Such collective behavior only for $\rho<\rho_c$ and not for $\rho>\rho_c$, which eventually leads to the non-trivial form of $A(\rho)$, is another manifestation of the mixed (or one-sided) nature of the jamming transition.

\begin{figure}
\onefigure[width=\columnwidth]{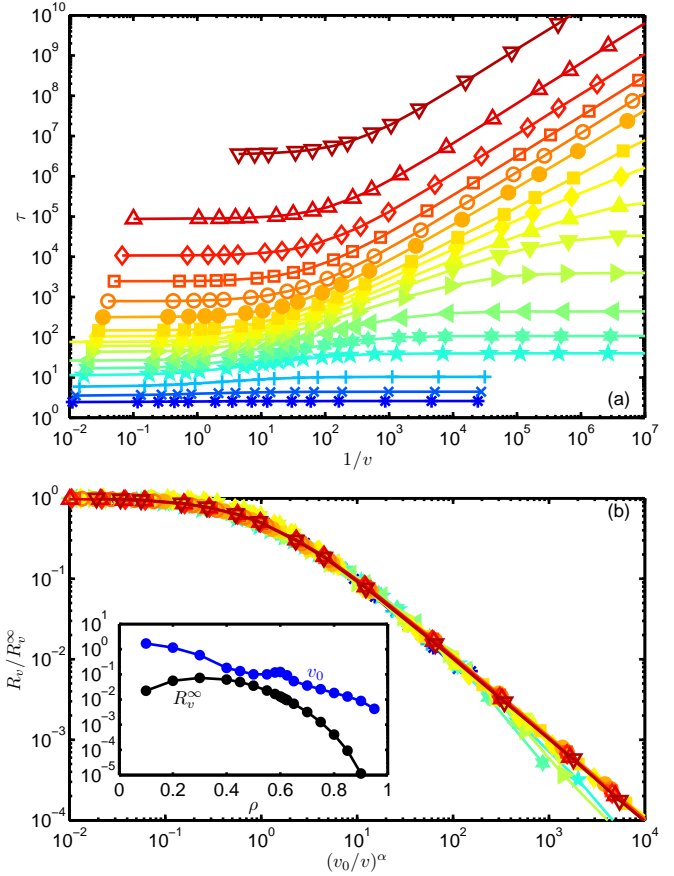}
\caption{(Color online) a) Relaxation time vs inverse driving for various densities (same legend as fig.~\ref{fig:T}). b) Collapse of relaxation rate due to the driving mechanism for all densities by scaling $R_v$ by the $v=\infty$ value and scaling by  $(v_0/v)^{\alpha}$. $R_v^{\infty}(\rho)$ and $v_0(\rho)$ are given in the inset and $\alpha(\rho)$ is given in fig.~\ref{fig:T}b.}
\label{fig:v}
\end{figure}

Figure~\ref{fig:v}a shows the relaxation time vs driving for various densities at zero temperature. We now write %
\begin{equation}
\frac{1}{\tau(\rho,v)} = R_{\rho}(\rho) + R_v(\rho,v),
\end{equation}
where $R_v$ is the relaxation rate due to the \emph{driving mechanism}, in which the neighborhood of a blocked particle changes due to driving events that unblock it and enable it to move subject to the kinetic constraint. We obtain $R_v$ by subtracting $R_{\rho} = 1/\tau(\rho,v=0)$ from $1/\tau(\rho,v)$ and show in fig.~\ref{fig:v}b that $R_v/R_v^{\infty}$ has the same dependence on $(v_0/v)^{\alpha}$ for all $\rho$.  Here, $R_v^{\infty}(\rho)$ describes the relaxation rate at infinite driving strength, $v_0(\rho)$ decreases monotonically with $\rho$, and $\alpha(\rho)$ behaves similarly to $A(\rho)$ (see insets to figs.~\ref{fig:T}b and \ref{fig:v}b).  Note that above $\rho_c$ and at densities below $\rho_c$ but away from the vicinity of the transition, $\alpha \approx 1$.  Thus, $R_v$ varies linearly with the flow velocity, $v$, at small $v$, as expected.  At higher $v$, for all densities, $R_v$ crosses over to a constant at $v \approx v_0$ because the neighborhood around a particle is completely randomized between attempts of diffusive moves.  As a result, increasing the driving strength even more does not affect the relaxation rate at high flow rates.  

\begin{figure}
\onefigure[width=\columnwidth]{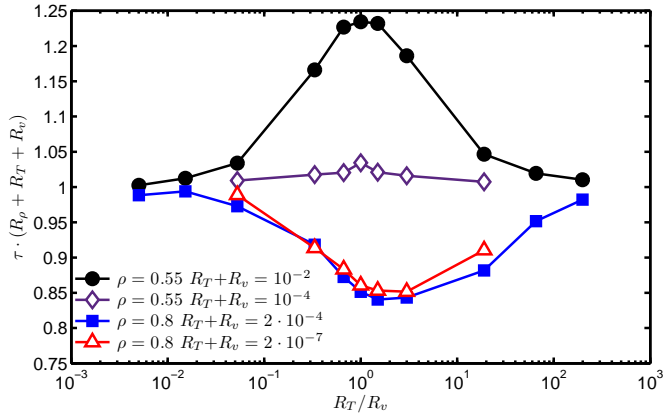}
\caption{(Color online) Ratio of the simulated relaxation time to the prediction of eq. (\ref{eq:TV}) vs the ratio of temperature to driving relaxation rates.}
\label{fig:TF}
\end{figure}

We now consider the interplay between temperature and driving. So far, we have identified the thermal relaxation rate, $R_T$, for $v=0$, and the driving relaxation rate, $R_v$, for $T=0$.  For $T>0$ and $v>0$, the simplest assumptions are that $R_T$ does not depend on $v$, $R_v$ does not depend on $T$, and the relaxation rates are additive, so that 
\begin{equation}
\tau(\rho,T,v) = \frac{1}{R_{\rho}(\rho)+R_T(\rho,T)+R_v(\rho,v)}. \label{eq:TV}
\end{equation}
Figure~\ref{fig:TF} shows the ratio of the actual relaxation time measured in simulations in which both $T>0$ and $v>0$ to this prediction. Obviously, when the values of $R_T$ and $R_v$ are very different, the smaller rate becomes irrelevant and the larger behaves as it behaves in the complete absence of the smaller. When $R_T$ and $R_v$ are comparable, deviations of around 20\% are seen, indicating that the two mechanisms are coupled and the relaxation is not given as a simple sum of independent relaxations. For $\rho<\rho_c$, these deviations disappear for small values of $R_T+R_v$ since then the density mechanism dominates.

\section{Dynamic Heterogeneity}

\begin{figure}
\onefigure[width=\columnwidth]{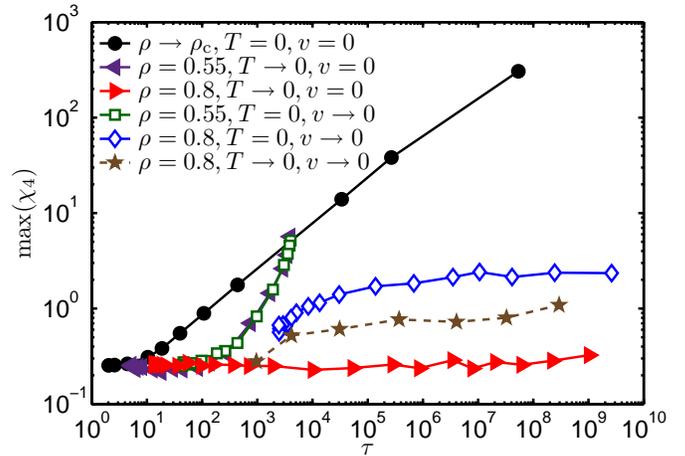}
\caption{(Color online) Dynamic heterogeneity vs relaxation time along various paths to jamming, as indicated in the legend.}
\label{fig:xi4}
\end{figure}

Differences between the relaxation mechanisms are clearly visible in spatial correlations of the dynamics~\cite{Glotzer,Franz,Dauchot,Keys}. For a given $\rho$, $T$, and $v$, $\chi_4(t)$ is maximal roughly when $t=\tau$ with a value related to the typical number of particles that rearrange collectively~\cite{AbateDurian2007,Toninelli_PRE_2005}.  Figure~\ref{fig:xi4} shows the maximal value of $\chi_4$ vs $\tau$ along different paths in the jamming phase-diagram. Solid circles denote the path of increasing density at $T=0$ and $v=0$.  As expected, along this path, the typical size of clusters rearranging collectively diverges as the system jams~\cite{ToninelliBiroliFisher2006}. The data behaves as $\rm{max}(\chi_4) \propto \tau^\lambda$ with $\lambda \approx 0.5$, in reasonable agreement with granular~\cite{AbateDurian2007} and colloidal~\cite{Brambilla} experiments.

When $T$ decreases at fixed density, $\chi_4$ does not diverge unless $\rho=\rho_c$.  For $\rho<\rho_c$ both the density and temperature relaxation mechanisms are at play.  A blocked particle can move by waiting until its neighbors move to unblock it via the density mechanism.  Additionally, temperature assists relaxation not only by allowing a blocked particle to overcome the kinetic constraint on its own, but also by allowing the neighbors of this blocked particle to overcome their kinetic constraints, thus releasing it and enabling it to move by a unblocked move.  Overall, the temperature mechanism is less collective than the density mechanism.  This is shown by the result in Fig.~\ref{fig:xi4} that $\chi_4$ is smaller (left-pointing triangles) at $T>0$ than along the $T=0$ path.  As $T$ decreases, the contribution of the temperature mechanism vanishes and the dynamics become dominated by the density mechanism and therefore become more collective with $\chi_4$ increasing until it meets the $T=0$, $v=0$ curve.  For $\rho>\rho_c$, relaxation occurs only via the temperature mechanism.  Such relaxation involves primarily a single blocked particle that waits a long time until it manages to move by a thermal move, and does not rely on the correlated dynamics of many particles.  Here, \emph{$\chi_4$ does not grow at all with decreasing temperature at $\rho>\rho_c$} (right-pointing triangles), indicating that the typical spatial size of each rearrangement does not increase as the dynamics slow down. 

Experimental data for many glass-forming liquids shows that the scale of heterogeneities is also essentially constant with decreasing temperature at sufficiently low temperatures.  There, an initial rise with $\tau$ in the number of molecules whose dynamics on the scale of  $\tau$ is correlated to a local enthalpy fluctuation is followed by a very slight increase or even saturation at large $\tau$~\cite{DalleFerrier}.   

When driving is lowered at fixed $\rho<\rho_c$ and $T=0$ (open squares in fig.~\ref{fig:xi4}), both the density and driving mechanisms cause relaxation, and as for the thermal case, $\chi_4$ grows until it meets the $T=0$, $v=0$ curve.  Interestingly, since the relative contribution of the density mechanism to the overall relaxation is what determines the heterogeneity along the $T \rightarrow 0$ and $v \rightarrow 0$ paths, it does not matter whether the non-collective rearrangement comes from temperature or from driving, and the curves for these two paths for $\rho<\rho_c$ superimpose in fig.~\ref{fig:xi4}.  For a path in which driving is lowered at $\rho>\rho_c$ and $T=0$, the density mechanism is frozen out and relaxation is due to driving alone.  As discussed earlier, the primary mechanism for relaxation in that case is that a blocked particle eventually becomes unblocked when its environment is changed by flow.  Like the temperature relaxation mechanism, this is a local process that does not become collective as the system jams.  Fig.~\ref{fig:xi4} shows that $\chi_4$ is slightly larger along the $v \rightarrow 0$ path than for the thermal case since more particles are involved in each rearrangement event, but the size of such events does not grow with increasing $\tau$ (open diamonds), as for the temperature mechanism.

Finally, we consider a trajectory at $\rho>\rho_c$ along which both temperature and driving are positive (stars in fig.~\ref{fig:xi4}).  To maximize the interplay between the temperature and driving mechanisms we select $T$ and $v$ such that $R_T=R_v$.  Since in this case the two non-collective mechanisms related to temperature and driving govern the relaxation dynamics, the spatial extent of dynamic heterogeneity saturates to a value which lies between that of the purely thermal trajectory (solid triangles) and that of the purely driven trajectory (open diamonds).

\section{Conclusions}

We studied a kinetically-constrained lattice-gas with a nontrivial jamming phase-diagram. Our model is substantially simpler than currently-used particulate models with more realistic interactions, and is easier to study numerically so that a wide range of time scales (over ten decades) may easily be probed even in relatively large systems.   

Our model introduces three mechanisms by which density fluctuations can relax, which we term the density, temperature and driving mechanisms.  In particulate systems the same physical mechanisms come into play, but they are intermingled in a more complicated way.  For example, real liquids, which are at high density, behave as if they have a lower effective density at high temperature.  This is because increasing temperature in a particulate system increases the ability to open up free volume, and hence effectively decreases the density~\cite{ning}.  It also decreases the effective particle diameter, which effectively decreases density~\cite{berthierwitten}.  Thus in real liquids, one would expect the effective density to increase as temperature decreases, thus slowing down relaxation due to the density mechanism.  Once the density mechanism becomes too slow and is frozen out, the relaxation time should be dominated by the temperature mechanism, where small numbers of particles overcome energy barriers via relatively uncorrelated rearrangements.  Thus, it seems reasonable to expect that the sharp crossover from the density mechanism to the temperature one at $\rho_c$ in our model should be replaced in more realistic systems by a gradual crossover from a density-dominated regime at low densities or high temperatures, to a temperature-dominated one at high density and low enough temperature.  More generally, at sufficiently high densities and low temperatures or driving, the density mechanism should eventually freeze out, leaving only the temperature and driving mechanisms to relax the system.  In that regime, our results suggest that the spatial scale of dynamic heterogeneities should be limited, as observed in glass-forming liquids~\cite{DalleFerrier}.  Recent analyses of other lattice-based models find similar results~\cite{Eisenberg,Harrowell}.

One corollary of this result is that one might expect the crossover from the density-mechanism-dominated regime at high temperature to the temperature-mechanism-dominated regime at low temperature to affect the form of the dynamics.  It is known for hard-sphere systems, where only the density mechanism is at play, that the increase of relaxation time with decreasing temperature at fixed pressure is super-Arrhenius (fragile behavior)~\cite{ning}.  On the other hand, the temperature mechanism should give rise to Arrhenius behavior (strong behavior).  Thus, the crossover from a regime controlled by the density mechanism to one controlled by the temperature mechanism might be accompanied by a fragile-to-strong crossover in the dynamics, although it is possible that one crossover is more abrupt than the other. 

\acknowledgments
We thank Jean-Louis Barrat, Eli Eisenberg, Tom Haxton, Claus Heussinger, Randy Kamien, Carl Modes, David Mukamel, Chris Santangelo, Reem Sari, Peter Sollich, Marija Vucelja, and Ning Xu for helpful discussions. This work is supported by NSF MRSEC grant DMR-0520020 and DE-FG02-05ER46199.


\begin{thebibliography}{0}
\bibitem{yield_stress_footnote}
For mechanical systems, stress is used as the non-equilibrium axis in the jamming phase-diagram, and there is a jammed region below the yield stress. In our model we set the flow rate and do not try to assign the stress required to generate it. Hence, the jammed phase in fig.~\ref{fig:T0F0}a reduces to a line.
\bibitem{footnote_px}
In order to correctly measure relaxation in the presence of driving, the persistence function is defined only in terms of motion along the vertical direction, perpendicular to the driving.
\end{thebibliography}
\end{document}